\documentclass{PoS}

\title{Central Exclusive Production at LHCb}

\ShortTitle{CEP at LHCb}

\author{\speaker{R.~McNulty} \\
        School of Physics, University College Dublin,\\
        Dublin 4, Ireland,\\
        (on behalf of the LHCb collaboration.)\\
        E-mail: \email{ronan.mcnulty@ucd.ie}}

\abstract{%

Central Exclusive Production (CEP) is a unique process at hadron machines in which
particles are produced via colourless propagators.
LHCb have measured the cross-sections for the CEP of vector mesons,
$J/\psi,\psi(2S),\Upsilon(1S),\Upsilon(2S)$ and $\Upsilon(3S)$, which are photo-produced.
In the double pomeron exchange process, preliminary measurements have been made
of $\chi_{c0},\chi_{c1}, \chi_{c2}$ meson production while the first observations of
the CEP of pairs of charmonia, $J/\psi J/\psi$ and $J/\psi \psi(2S)$, have been made
and limits obtained on the pair production of other charmonia. 
}

\FullConference{XXIV International Workshop on Deep-Inelastic Scattering and Related Subjects\\
		11-15 April, 2016\\
		DESY Hamburg, Germany}

\begin{document}

%

\section{Introduction}
Central exclusive production (CEP) at the LHC is characterised by an isolated system of particles
surrounded by two rapidity gaps that extend  down to the intact colliding protons; 
see \cite{review} and the papers referenced in \cite{ijmpa} for a recent survey of the field.
The lack of additional activity signals the presence of colourless propagators: two photons; two pomerons;
or a photon and a pomeron.
Measurements of CEP test QCD, investigate the nature of the pomeron,
and can be used to constrain the gluon PDF.  At $\sqrt{s}=13$ TeV, measurements of $J/\psi$
production will probe the gluon PDF down to a fractional momentum of the proton
$x=2\times10^{-6}$, a scale at which saturation effects may
become visible.

The LHCb detector~\cite{lhcb} 
is suited to studying CEP as it is fully instrumented with tracking, calorimetry and 
particle identification in the pseudorapidity, $\eta$, range between 2 and 5.  In addition, charged activity in the
backward region, in the approximate range $-3.5<\eta<-1.5$, can be vetoed
due to the presence of a silicon strip detector that surrounds the 
interaction point.
LHCb is designed to trigger on particles produced at low transverse momentum. 
CEP events are selected by triggering on muons with transverse momentum, $p_T$, above 400 MeV/c or electromagnetic or hadronic energy above 1000 MeV, in coincidence with a 
total event charged multiplicity of 
less than 10 as recorded by a scintillating pad detector.
A further advantage of LHCb for CEP is the low number of proton-proton interactions (typically 1.5)
per beam crossing.

The LHCb measurements of CEP to date have concentrated on final states with muons.  
In this report, I focus on measurements of the photoproduction of single charmonium and bottomium 
as well as the production of
double charmonia, which is principally produced by double pomeron exchange. 
The single charmonium measurements use almost 1 fb$^{-1}$ of data taken at $\sqrt{s}=7$ TeV, while
the other measurements add an additional 2 fb$^{-1}$ at $\sqrt{s}=8$ TeV.  
Preliminary results, based on 37 pb$^{-1}$ of data, are also available on $\chi_c$ production and dimuons produced 
by the QED diphoton exchange process~\cite{prelim}.
New forward scintillators have been installed for the $\sqrt{s}=13$ TeV running, and their impact on future measurements will be briefly discussed.

\section{Photoproduction of $J/\psi$ and $\psi(2S)$ mesons}

\begin{figure}[htb]
\centering
\includegraphics[width=7.5cm]{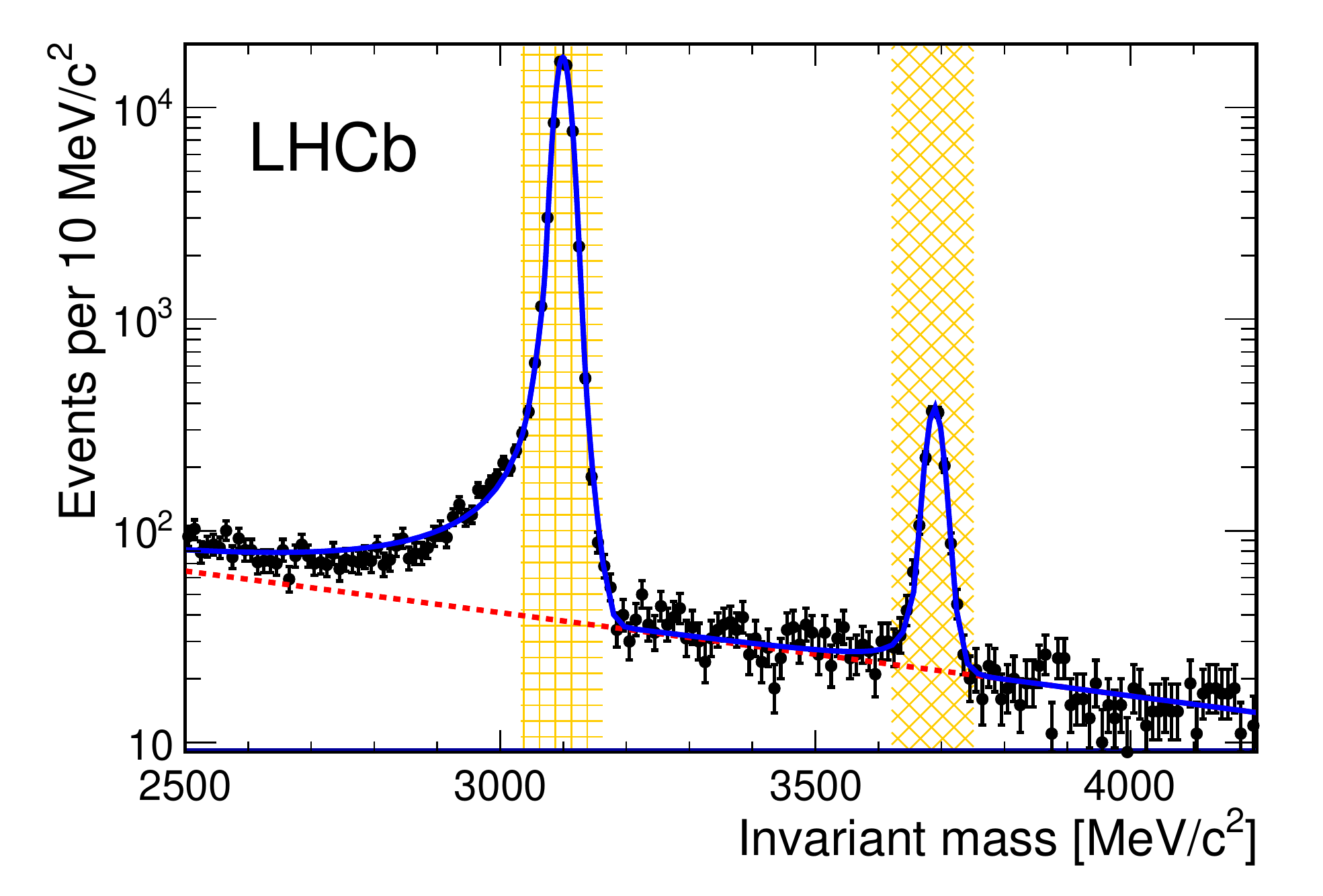}
\includegraphics[width=7.5cm]{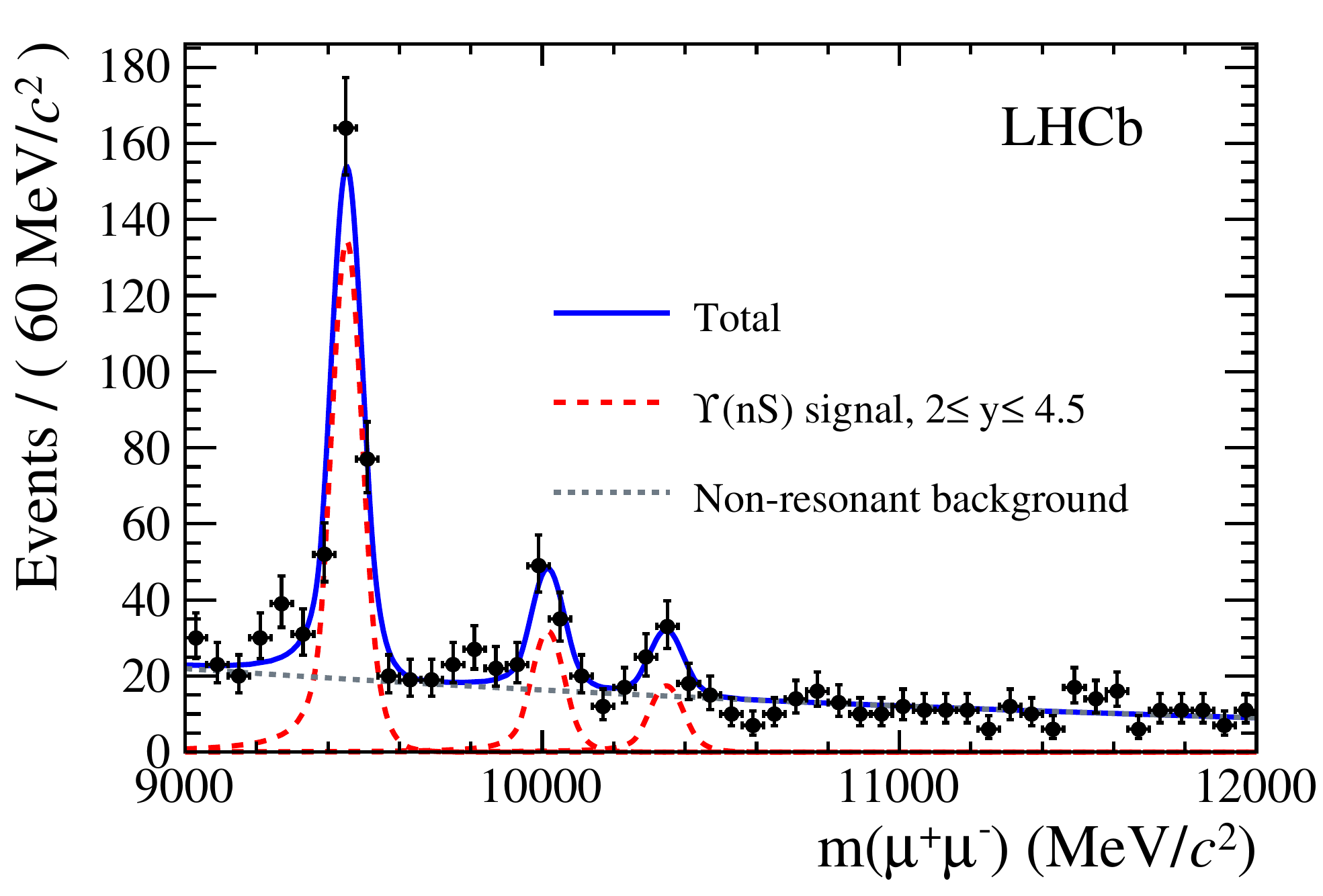}
\caption{Invariant mass of dimuons in 
(left) the charmonium analysis~\cite{lhcb_jpsi} with the $J/\psi$ and $\psi(2S)$ windows indicated and
(right) the bottomonium analysis~\cite{lhcb_ups}.}
\label{fig:mass}
\end{figure}

Candidates for $J/\psi$ mesons produced through CEP are selected~\cite{lhcb_jpsi} by requiring two
identified muons inside the LHCb acceptance and no photons or additional tracks in either forward
or backward directions.  The $p_T^2$ of the dimuon is required to
be below 0.8 GeV$^2$/c$^2$, and its invariant mass to be within 65 MeV/c$^2$ of the known  $J/\psi$ or
$\psi(2S)$ masses.  The invariant mass of all candidates (with the mass requirement removed) is shown
in Fig.\ref{fig:mass} (left).  The non-resonant background, due to the QED production of dimuons via
photon propagators, is modelled with an exponential function and is estimated to account for
$(0.8\pm0.1)$\% of the $J/\psi$ 
and $(16\pm 3)$\% of the $\psi(2S)$ sample.
Feed-down backgrounds inside the $J/\psi$ mass window, 
amounting to $(10.1\pm 0.9)$\%, 
are due to $\chi_c$
or $\psi(2S)$ mesons decaying to $J/\psi$ and photons that are undetected due to being very soft or going
outside the detector acceptance.  
Inelastic $J/\psi$ or $\psi(2S)$ production, in which the proton dissociates but does not produce
activity inside the LHCb acceptance, is assessed by fitting the $t\approx p_T^2$ distribution and assuming
that $d\sigma/dt$ can be modelled by two exponentials for signal and background, as assumed in
Regge theory and observed at HERA~\cite{h1,h1p}.  
The fitted parameters for the exponentials are consistent with
those found at HERA, having corrected for kinematic differences.
In total, $(59.2\pm 1.2)$\% of the $J/\psi$ sample and $(52\pm7)$\% of the $\psi(2S)$ sample is estimated to be exclusively produced.
 
\begin{figure}[htb]
\centering
\includegraphics[width=6.5cm]{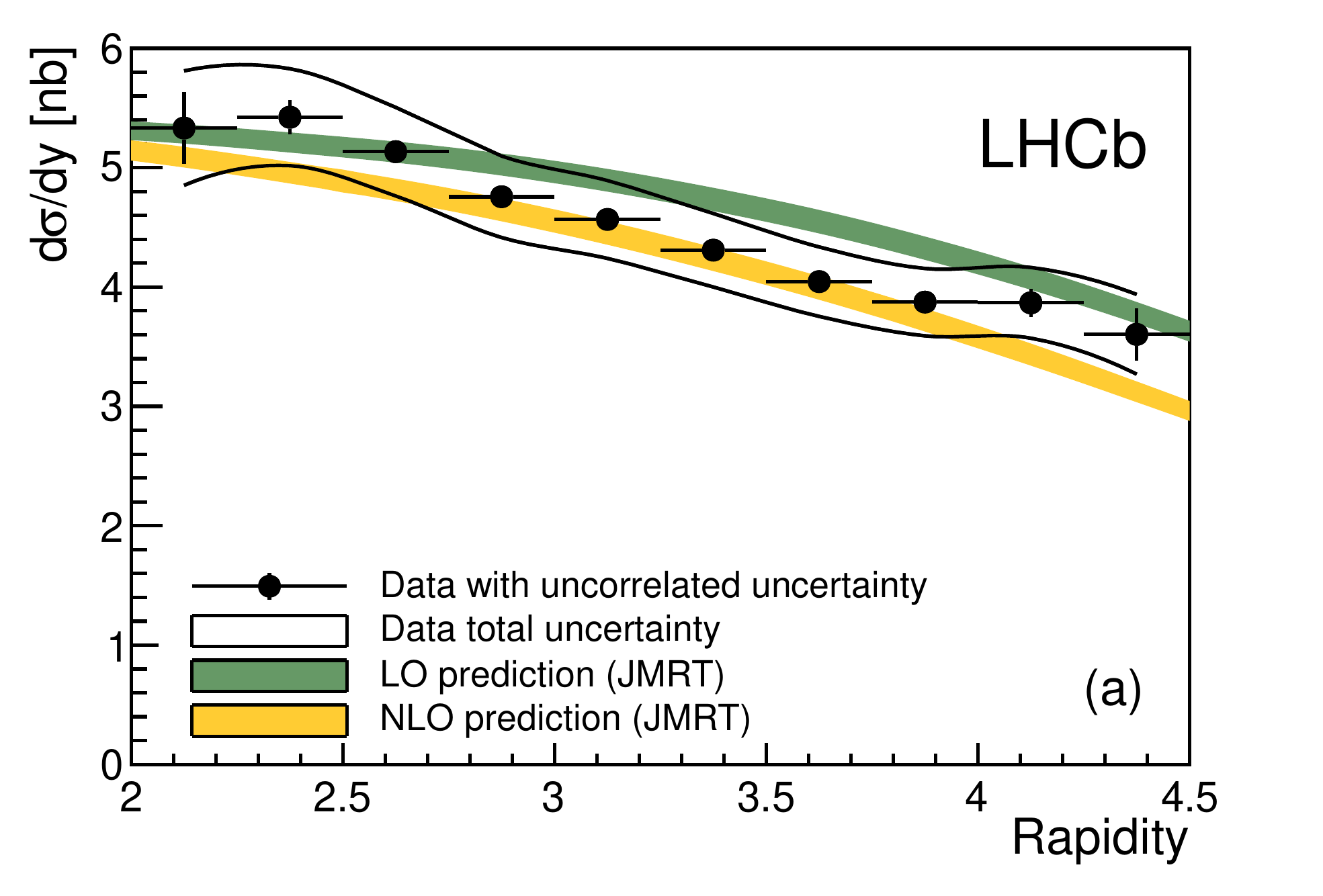}
\includegraphics[width=6.5cm]{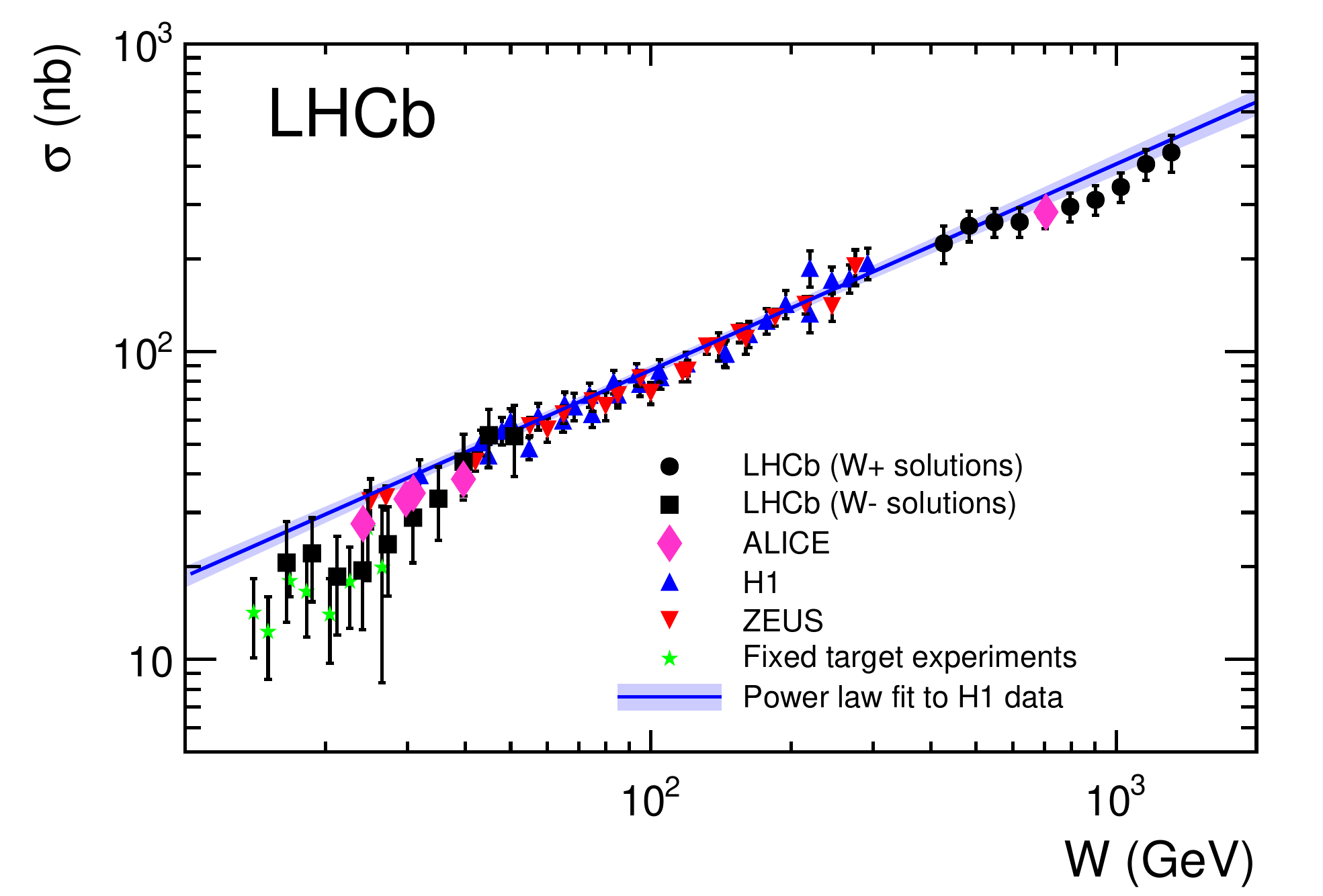}

\includegraphics[width=6.5cm]{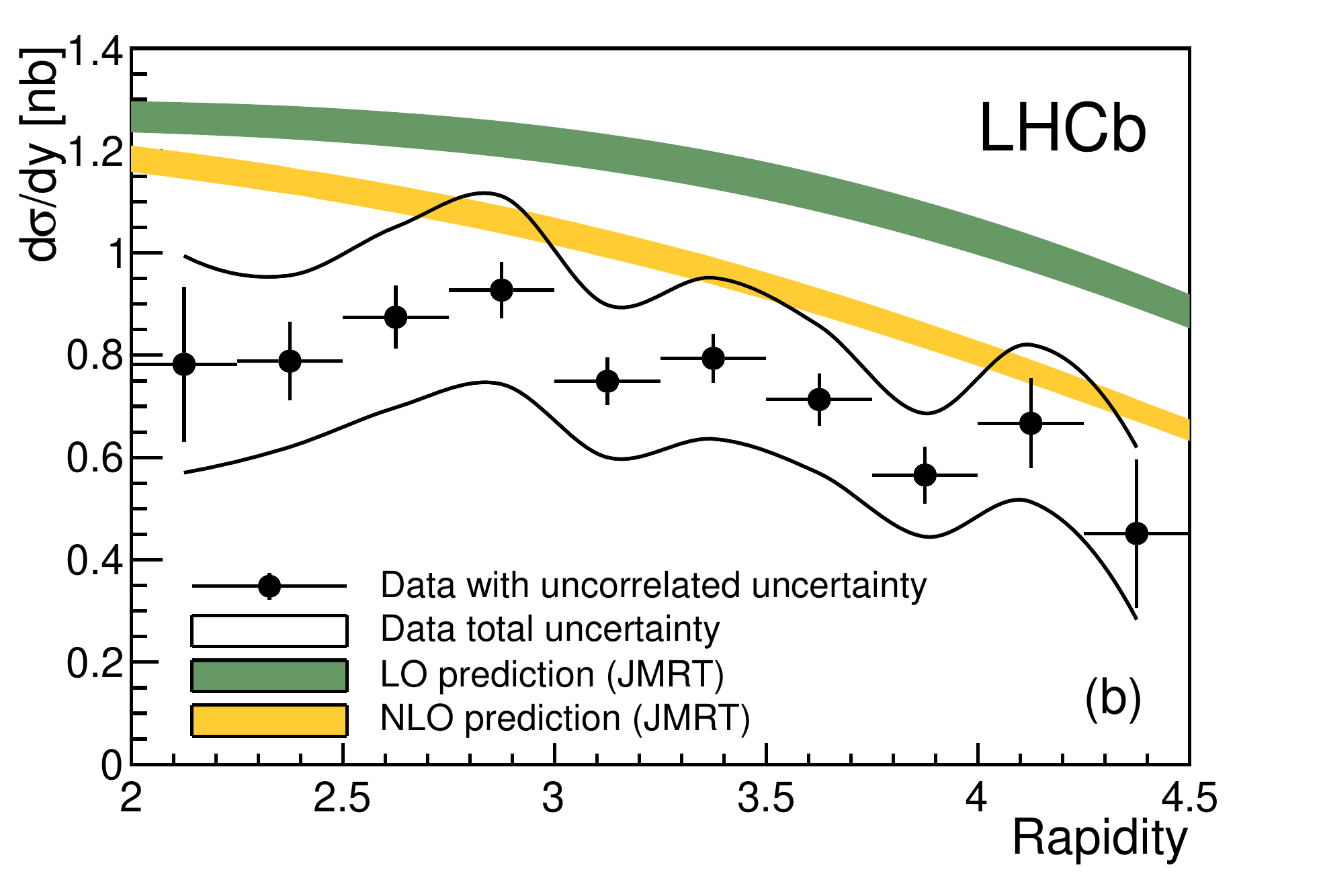}
\includegraphics[width=6.5cm]{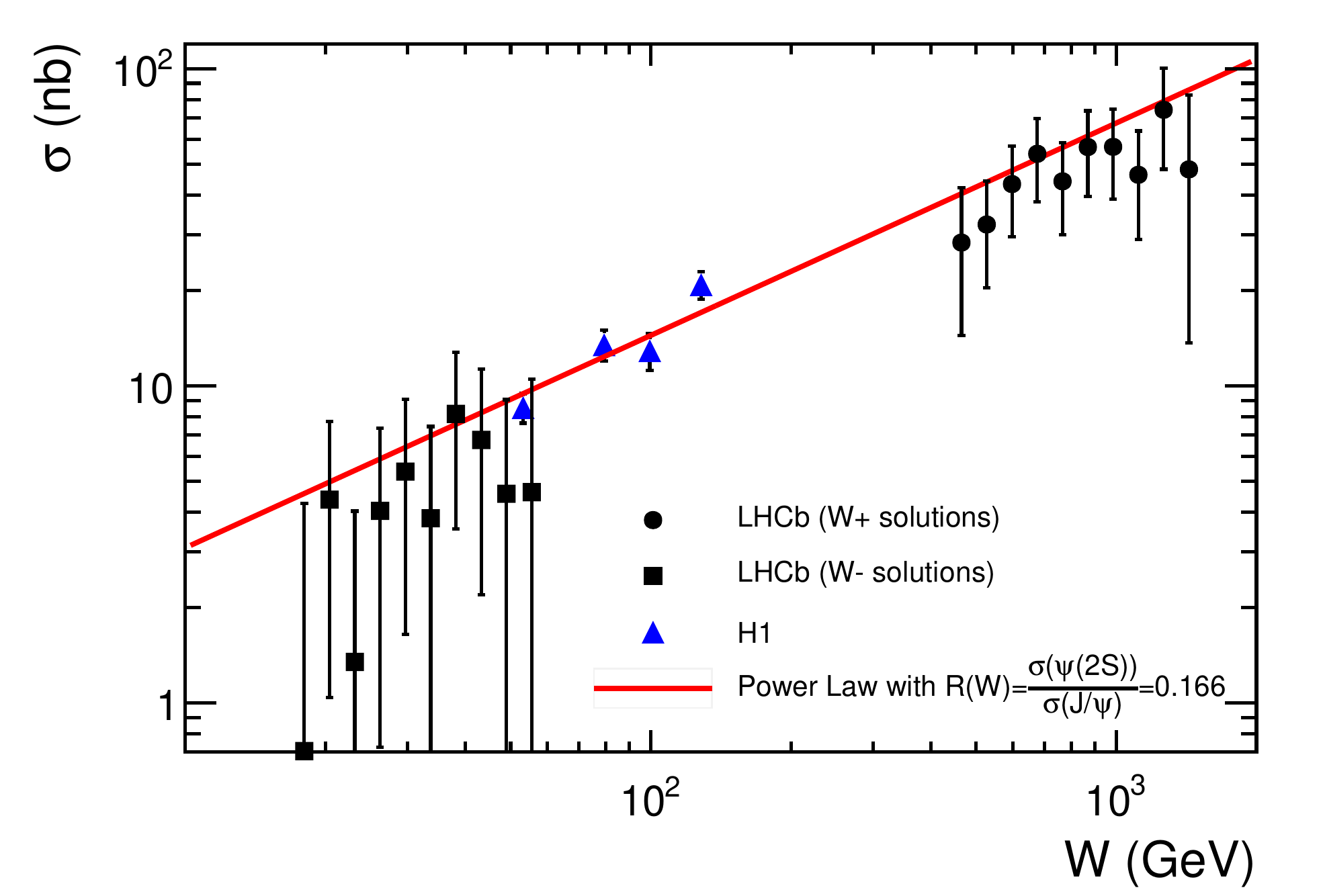}

\includegraphics[width=6.5cm]{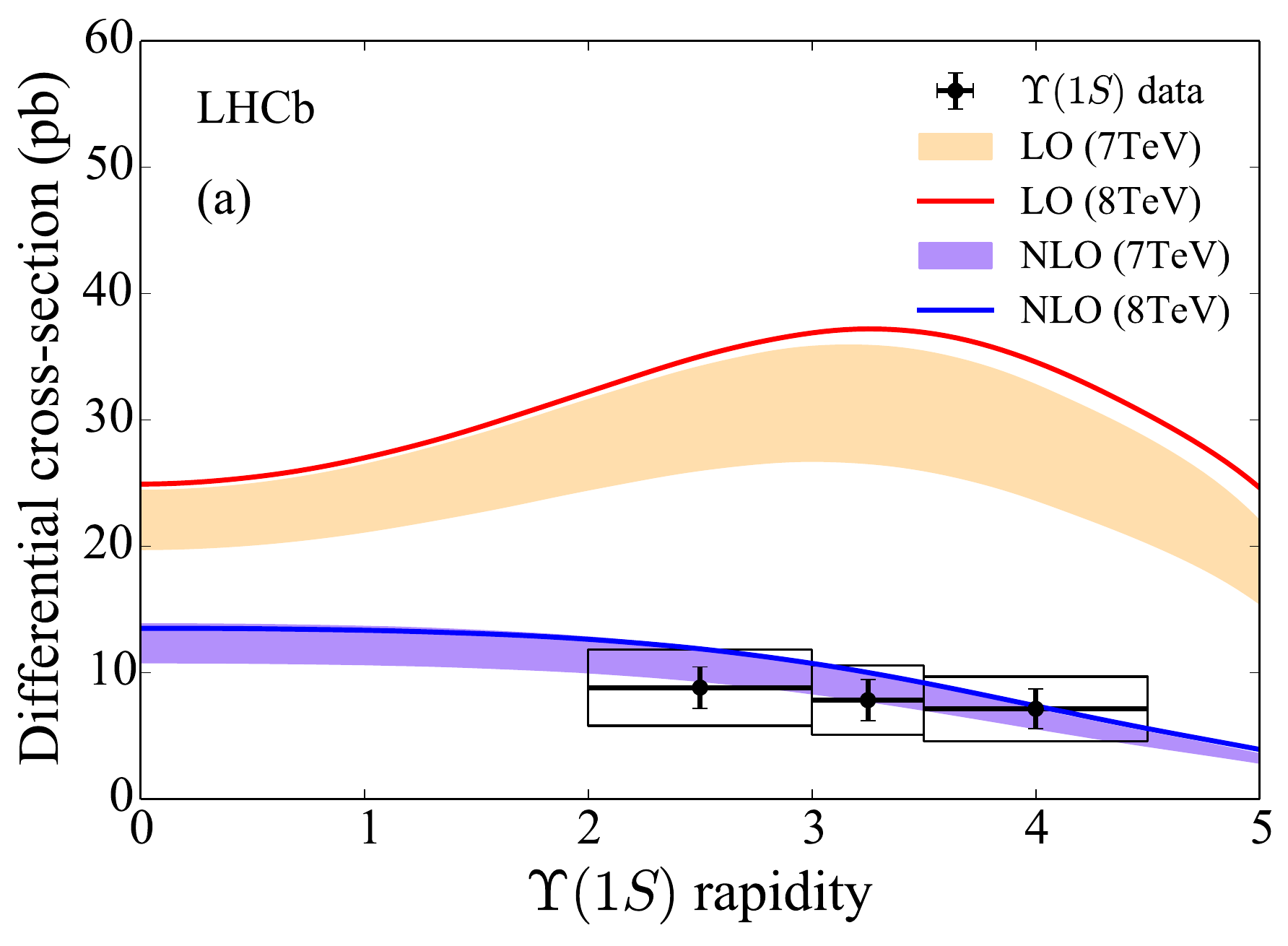}
\includegraphics[width=6.5cm]{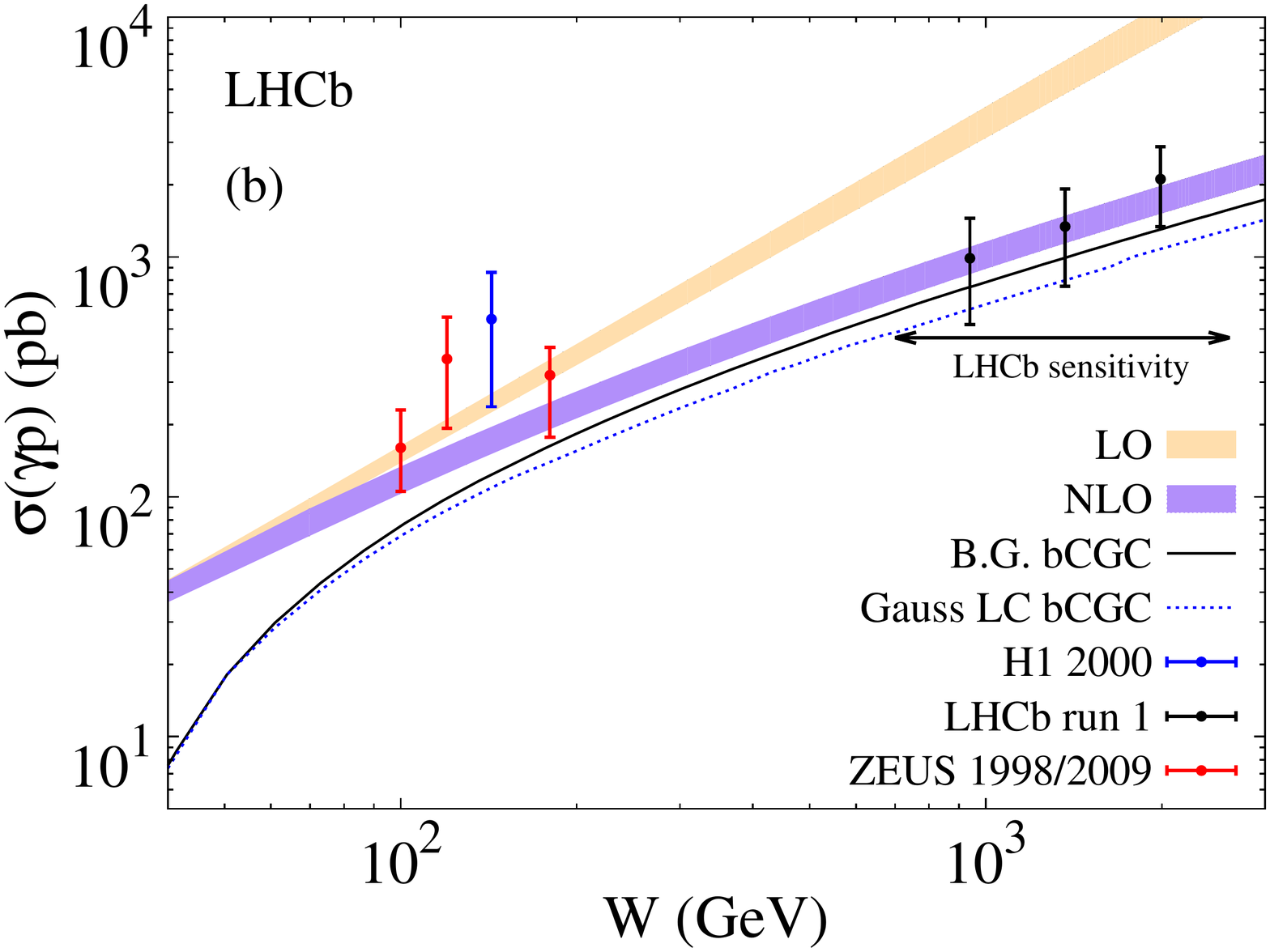}

\caption{
Differential cross-sections for exclusively produced 
(top left) $J/\psi$,
\mbox{(centre left) $\psi(2S)$}, and
(bottom left) $\Upsilon$,
compared to LO and NLO predictions~\cite{jmrt}.
Compilation of photoproduction cross-sections from various experiments for
(top right) $J/\psi$,
(centre right) $\psi(2S)$,
(bottom right) $\Upsilon$.
The LHCb results are model-dependent as explained in the text.
}
\label{fig:cs}
\end{figure}

The cross-section for CEP charmonia is determined from the estimated number of CEP events, 
correcting for the detector efficiency and acceptance: 
the former is determined with tag-and-probe techniques in data while the latter is found from
simulated events.
The total cross-sections are given in Table~\ref{tab:cs} while differential cross-sections
as a function of rapidity are shown in Fig.~\ref{fig:cs} compared to LO and
approximate NLO predictions from \cite{jmrt}.  Results on the total cross-section have
been compared to other predictions ~\cite{gm,mw,ss,star,super} and all agree with the data.
A photoproduction cross-section can be derived from these results once 
rapidity gap corrections and photon flux factors are included.  A two-fold ambiguity is present
due to not knowing which of the protons the photon was radiated from.  
This can be resolved in a model-dependent way by assuming the H1 derived 
power-law~\footnote{The power-law for $J/\psi$ is taken from \cite{h1} while for $\psi(2S)$,
I use $R(W)=\sigma(\psi(2S))/\sigma(J/\psi)=0.166$ from \cite{h1p}.}
for one of the solutions.
The (model-dependent) cross-section derived is shown in 
Fig.~\ref{fig:cs}, compared to results from HERA, fixed target collisions, and 
proton-lead collisions at ALICE~\cite{alice}, in which the aforementioned ambiguity can be resolved. 

\begin{table}[htb]
\begin{tabular}{|c|c|c|}
\hline
Quantity measured & Kinematic region & Measurement (pb)\\
\hline
$\sigma(pp\rightarrow p J/\psi p) \cdot BR(J/\psi\rightarrow \mu\mu)$  &
$2<\eta_\mu,y_{J/\psi}<4.5$& 
$291\pm 7 \pm 19$~\cite{lhcb_jpsi} \\
$\sigma(pp\rightarrow p \psi(2S) p) \cdot BR(\psi(2S)\rightarrow \mu\mu)$  & 
$2<\eta_\mu,y_{\psi(2S)}<4.5$& 
$6.5\pm 0.9\pm 0.4$~\cite{lhcb_jpsi} \\
$\sigma(pp\rightarrow p \Upsilon(1S) p) $  &
$2<\eta_\mu,y_{\Upsilon(1S)}<4.5$&  
$9.0\pm2.1\pm 1.7$~\cite{lhcb_ups} \\
$\sigma(pp\rightarrow p \Upsilon(2S) p) $  & 
$2<\eta_\mu,y_{\Upsilon(2S)}<4.5$& 
$1.3\pm0.8\pm0.3$~\cite{lhcb_ups} \\
$\sigma(pp\rightarrow p \Upsilon(3S) p) $  &
$2<\eta_\mu,y_{\Upsilon(3S)}<4.5$&  
$<3.4$ at 95\% c.l.~\cite{lhcb_ups} \\
$\sigma(J/\psi J/\psi ) $  &
$2<y_{J/\psi J/\psi}<4.5$&  
 $58\pm 10\pm 6$~\cite{djpsi}\\
$\sigma(J/\psi \psi(2S)) $  &
$2<y_{J/\psi\psi(2S)}<4.5$&  
$63^{+27}_{-18}\pm 10$~\cite{djpsi} \\
$\sigma( \psi(2S)\psi(2S) $  &
$2<y_{\psi(2S)\psi(2S)}<4.5$&   
$<237$ at 90\% c.l.~\cite{djpsi} \\
$\sigma(\chi_{c0}\chi_{c0})$ &
$2<y_{\chi_{c0}\chi_{c0}}<4.5$&   
$<69000$ at 90\% c.l.~\cite{djpsi} \\
$\sigma(\chi_{c1}\chi_{c1})$ &
$2<y_{\chi_{c1}\chi_{c1}}<4.5$&   
$<45$ at 90\% c.l.~\cite{djpsi} \\
$\sigma(\chi_{c2}\chi_{c2})$ &
$2<y_{\chi_{c2}\chi_{c2}}<4.5$&   
$<141$ at 90\% c.l.~\cite{djpsi} \\

\hline
\end{tabular}
\caption{Total cross-section results for charmonia and bottomonia states.
The double charmonia cross-sections include events where proton dissociation occurs.} 
\label{tab:cs}
\end{table}

\section{Photoproduction of $\Upsilon$ mesons}
Candidates for $\Upsilon$ mesons produced through CEP are selected~\cite{lhcb_ups} 
by requiring two muons
in LHCb and no other charged tracks.  The muon pair must have $p_T^2<2$ GeV$^2$/c$^2$ and an invariant mass between 9 and 20 GeV/c$^2$, which allows the shape of the continuum to be
determined at the same time as the fit to the $\Upsilon(1S),\Upsilon(2S)$ and $\Upsilon(3S)$.   
The distribution of candidates is shown in Fig.\ref{fig:mass} (right).

Feed-down backgrounds coming from the various $\chi_b$ states are estimated
to contribute $39\pm7$\% of the total signal yield.
As in the charmonium analysis, inelastic $\Upsilon$  production is assessed by fitting the $t\approx p_T^2$ distribution, but with the signal shape given by the {\tt SUPERCHIC} generator\cite{super}.  
After subtraction of the $\chi_b$ component, $(54\pm 11)$\% is assessed to be exclusively produced;
thus of the total $\Upsilon$ yield, one third is CEP.

After correcting for the detection efficiency, found using simulated events, 
and taking account of the luminosity, 
the cross-sections are determined and reported in Table~\ref{tab:cs}.
The low statistics and sizeable background impact on the significance of the 2S and 3S states.
However, there are sufficient statistics to divide the 1S state into three bins of rapidity, which are
plotted in Fig.\ref{fig:cs}.  Compared to the charmonium analysis, the difference between the
LO and NLO predictions is large and the data show a clear preference for the latter.
A photoproduction cross-section can be derived from these results.  Here, the smaller of the two
ambiguous solutions (that contributes between 5\% and 20\%) is ignored.  The result is shown in
Fig.\ref{fig:cs} showing good consistency with HERA results~\cite{h1_ups,z_ups}.

\section{Production of pairs of charmonia}

The pair production of charmonia has been observed inclusively in a previous LHCb analysis~\cite{lhcb_in}, where tetraquark contributions or
double parton scattering (DPS) may play an important role.  The contribution of DPS in CEP
is minimal and the process proceeds dominantly through double pomeron exchange. 
The cross-section for CEP of $J/\psi J/\psi$ inside the LHCb acceptance is predicted to lie in the range 2-20 pb,
depending on the model used for the soft survival factor and the gluon PDF that enters with the 
fourth power~\cite{djpsi_th}.

\begin{figure}[htb]
\centering
\includegraphics[width=7.5cm]{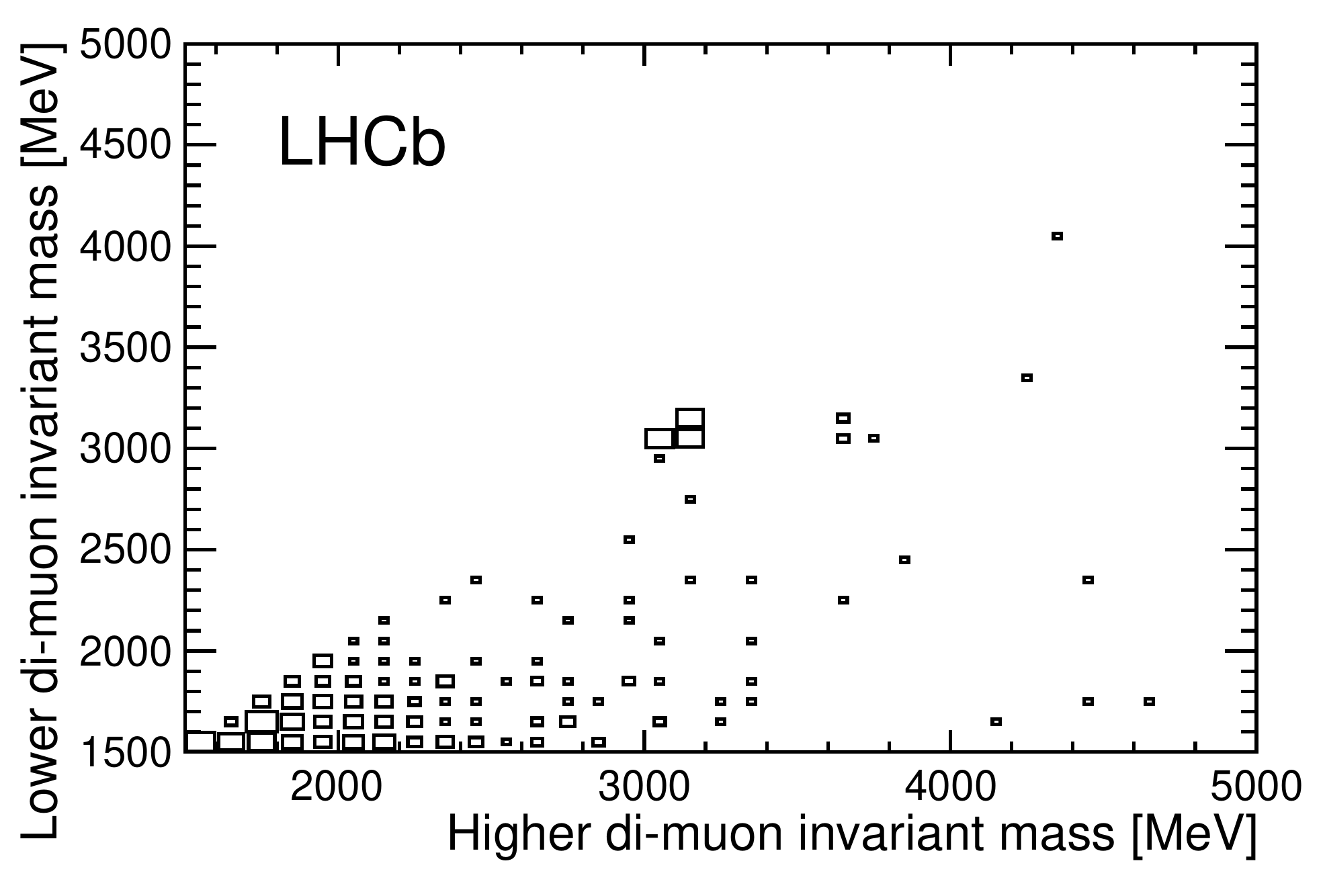}
\includegraphics[width=7.5cm]{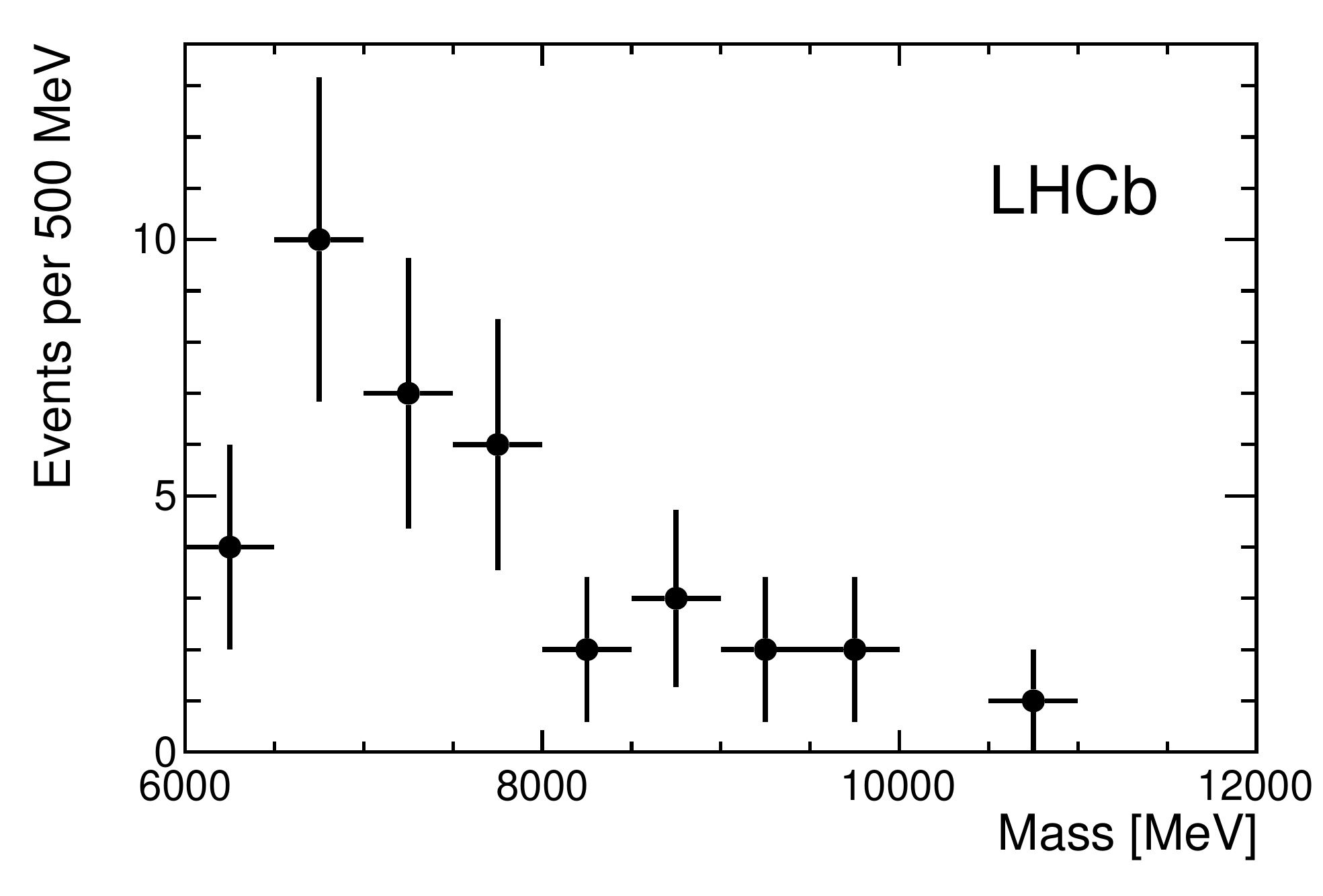}

\caption{Invariant mass of
(a, left) dimuon pairs
(b, right) the $J/\psi J/\psi$ system~\cite{djpsi}.
}
\label{fig:dj}
\end{figure}

The selection of the CEP of charmonia pairs~\cite{djpsi} requires exactly four tracks, at least three of which are identified
as muons.  The invariant masses of two unlike-sign pairs of muons is required to be consistent with
that of the $J/\psi$ or $\psi(2S)$ mesons.  The pairwise combinations are shown in 
Fig.\ref{fig:dj}(a); 37 candidates are consistent with $J/\psi J/\psi$ production and 5 are consistent
with $J/\psi \psi(2S)$.  
The invariant mass of the four tracks is shown in Fig.\ref{fig:dj}(b) and agrees qualitatively with
the spectrum observed inclusively~\cite{lhcb_in}.
The signal is only seen when there are precisely 4 tracks in the event; it is not present when additional tracks are present in the detector.
After accounting for efficiency and luminosity, the cross-section for pairs of charmonia produced
in the absence of other charged or neutral activity in LHCb is determined and given in Table~\ref{tab:cs}.

Because of the low statistics, it is difficult to assess how often proton dissociation occurs.  A fit to the
$t$ distribution suggests $(42\pm 13)$\% of the sample is CEP implying an exclusive
cross-section for $J/\psi J/\psi$  
inside the LHCb acceptance of $(24\pm 9)$ pb, in broad agreement with the predictions. 

The analysis is also sensitive to the production of pairs of $\chi_c$ mesons, which
would be visible through their decays to a $J/\psi$ meson and a photon.
Searches are made for additional photons in events with two $J/\psi$ candidates.  A single event
is consistent with $\chi_{c0}\chi_{c0}$ production but is also consistent with partially reconstructed
$J/\psi \psi(2S)$ production.  Limits are set on pairs of tensor meson charmonia and are given
in Table~\ref{tab:cs}.

\section{Future prospects}

\begin{figure}[htb]
\centering
\includegraphics[width=7.5cm]{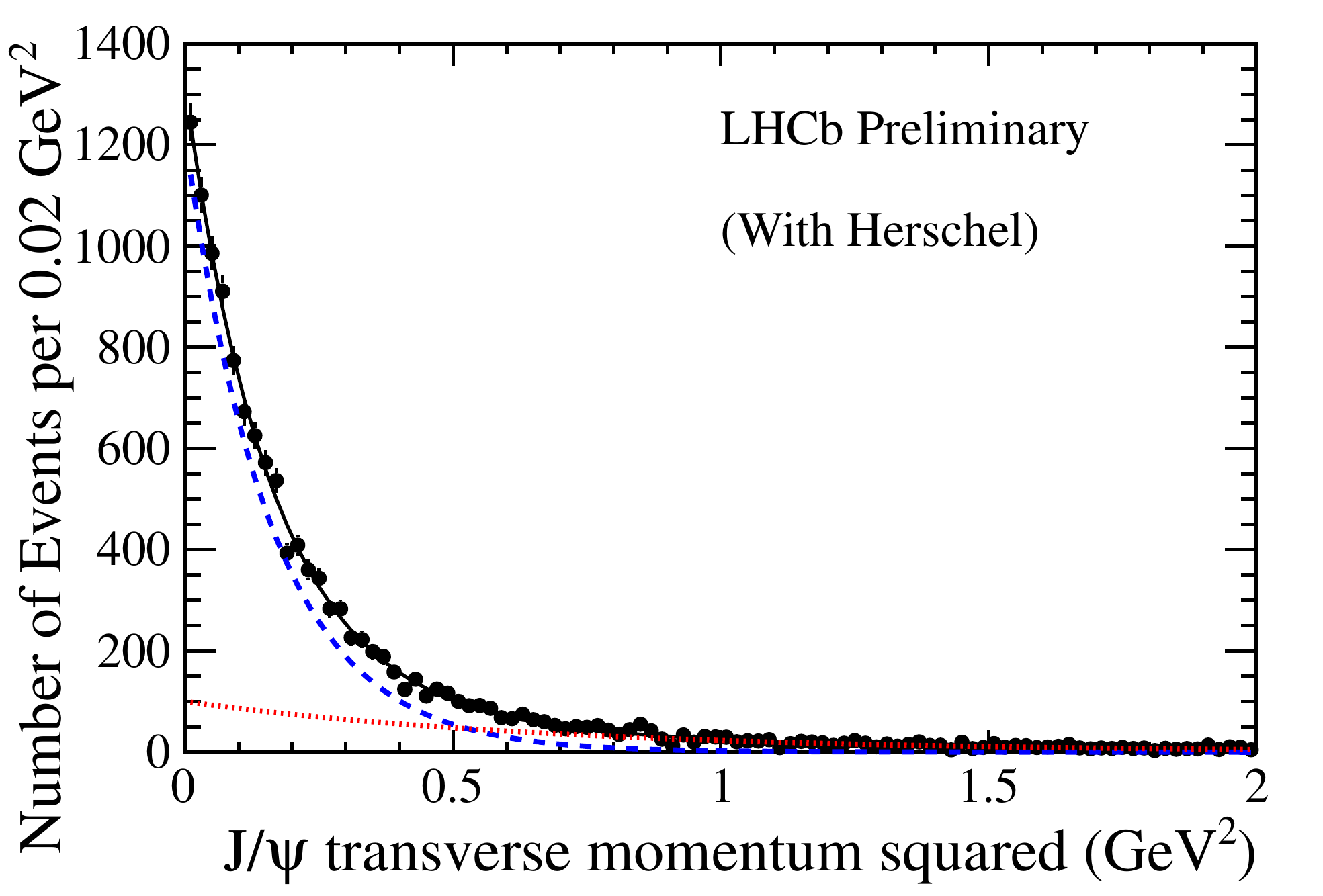}
\includegraphics[width=7.5cm]{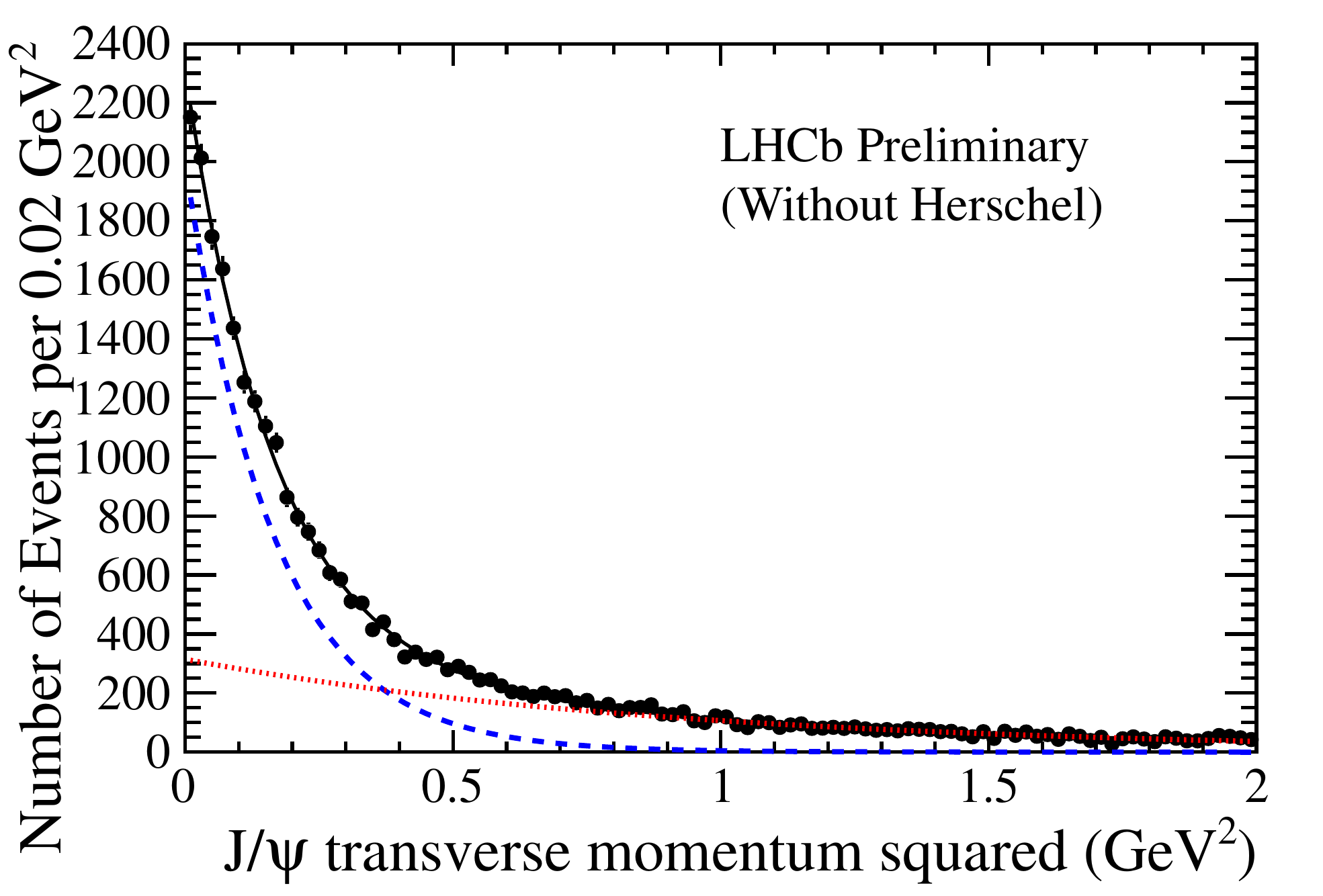}

\caption{Transverse momentum squared of $J/\psi$ candidates at $\sqrt{s}=13$ TeV
with (left) and without (right) the use of newly installed scintillators close to the beamline~\cite{13tev}.}

\label{fig:herschel}
\end{figure}

Experimentally, LHCb is sensitive to about 5.5 units in pseudorapidity allowing modest rapidity gap sizes to be 
identified.  The identification of CEP events would be significantly aided if the veto region could be
increased.  To this end, for the $\sqrt{s}=13$ TeV data-taking, forward shower counters, consisting of five planes of scintillators, have been installed 
perpendicular to the beam
at $-114$ m, $-19.7$ m, $-7.5$ m, $+20$ m and $+114$ m
from the interaction point. 
With these, LHCb has sensitivity to particles
in the regions $-10<\eta<-5, -3.5<\eta<-1.5, 1.5<\eta<10$.
The effect of the additional veto on the composition of a sample of $J/\psi$ candidates
gathered at $\sqrt{s}=13$ TeV can be seen in Fig.~\ref{fig:herschel}.
The estimated proton dissociation contribution is roughly halved in the signal region 
below $p_T^2=0.8$ GeV$^2$, with an even greater suppression 
at higher values.

Additional improvements for the $\sqrt{s}=13$ TeV running include a reduction 
in the hadronic trigger thresholds, which  will allow LHCb  
to extend current studies to investigate CEP of
light vector mesons, a region of spectroscopic interest where the presumed dominance of the
double pomeron exchange mechanism provides a gluon-rich laboratory to 
search for glueballs and saturation effects.

\end{document}